# Quantum key distribution over 122 km of standard telecom fiber

C. Gobby,[a] Z. L. Yuan, and A. J. Shields

Toshiba Research Europe Ltd, Cambridge Research Laboratory, 260 Cambridge Science Park,

Milton Road, Cambridge, CB4 0WE, UK

We report the first demonstration of quantum key distribution over a standard telecom fiber exceeding 100 km in length. Through careful optimisation of the interferometer and single photon detector, we achieve a quantum bit error ratio of 8.9% for a 122km link, allowing a secure shared key to be formed after error correction and privacy amplification. Key formation rates of up to 1.9 kbit/sec are achieved depending upon fiber length. We discuss the factors limiting the maximum fiber length in quantum cryptography.

03.67.Dd Quantum Cryptography

---

[a] Also with University of Cambridge, Department of Physics, Cavendish Laboratory, Madingley Road, Cambridge CB3 0HE, UK

There is currently much interest in using quantum cryptography as a method for distributing cryptographic keys on fiber optic networks,[1] the secrecy of which can be guaranteed by the laws of quantum mechanics. Since any information gained by an eavesdropper causes errors in the formed key, the sender (Alice) and receiver (Bob) can test its secrecy. Provided the quantum bit error rate (QBER) is less than a certain threshold,[2] error correction[3,4] and privacy amplification[5] can be applied to form a perfectly shared key with minimal information known to an eavesdropper.

Since the original proposal by Bennett and Brassard[6] in 1984, quantum key distribution (QKD) systems based upon transmission of encoded weak coherent pulses have been implemented by a number of groups.[7-12] The most successful approach for fiber optic based systems has been to encode the qubit information upon the phase delay in an interferometer. In the absence of a 'quantum repeater' which could regenerate a modulated photon, photon loss in the fiber limits the maximum distance over which quantum cryptography may be applied. As the fiber length increases, the signal rate falls to a value approaching that of the intrinsic error rate of the receiver's equipment. Eventually this results in the QBER exceeding the threshold for privacy amplification, preventing a secure key from being formed. Until now this has restricted demonstrations of QKD to fiber lengths shorter than 100 km. We show here that this barrier can be exceeded by minimising the contribution to the intrinsic error rate of detector dark count noise and stray photons. With some improvements distances up to 165 km could be possible.

Our system is based upon a time division Mach-Zender interferometer using phase modulation, as shown in Fig. 1. Polarising beam combiner and splitter are used so that photons from Alice's short arm are directed into Bob's long arm (S-L) and vice versa (L-S). The lengths of two routes S-L and L-S are roughly matched using a variable delay line in Alice's setup, with fine adjustment achieved by using a fiber stretcher in Bob's long arm. Thus the relative phase delays introduced to the two paths by Alice and Bob using their phase modulators, determine the probability that a photon exits either output of Bob's interferometer.

Photons are generated by a 1.55 μm distributed feedback pulsed laser diode operating at 2 MHz with a pulse width of 80 ps. The pulses are strongly attenuated so that on average 0.1 photons per clock cycle leaves Alice's apparatus. The intensity ratio of the reference pulse (though Alice's long arm) to the encoded pulse (through Alice's short arm) is 1.6:1, so that the encoded signal contains 0.04 photons per pulse on average. Phase modulators controlled by custom electronics, in the two interfering routes are used to encode the bit information. The signal is multiplexed with pulses from a 1.3 μm clock laser which serves as a timing reference. InGaAs avalanche photodiodes operating in gated mode, with a gate width of 3.5 ns and an excess voltage of 2.5 V, and cooled to an approximate temperature of -100 °C, are used to detect the single photons. It is imperative for operation over long fibers that the dark count rate in the single photon detector is as low as possible. Our detectors typically have a dark count probability of $10^{-7}$ per ns, along with a detection efficiency of around 12% at 1.55 μm. This corresponds to a Noise Equivalent Power of $1.1 \times 10^{-17}$ WHz$^{-1/2}$, which is one of the lowest reported to date at this operating temperature.

Thanks to careful alignment of the polarisation maintaining optics, the interferometer shows nearly perfect classical interference. An interference fringe visibility of 99.96% has been obtained for a 122km fiber link, as shown in the inset of Fig. 2. To our knowledge, this is the best visibility achieved so far for a quantum cryptographic interferometer. For quantum interference experiments, the signal laser was attenuated to 0.1 photons per clock cycle leaving Alice's apparatus. Bob's detector was synchronised by the 1.3 μm clock laser. The phase was varied by applying a DC bias to the piezo-driven fiber stretcher in the long arm of Bob's interferometer. Figure 2 shows the quantum interference visibility as a function of the fiber length, displaying values as high as 99.7% for short fiber lengths.

The quantum interference fringe visibility is greater than 99% for lengths up to 65 km and decreases for longer fiber lengths. Beyond 65 km, the fiber attenuation reduces the signal rate to a value comparable to the intrinsic error rate in Bob's detector. Nevertheless, by minimising the intrinsic error rate, we have achieved a visibility of 88.4% at 122 km. There are two major contributions to the intrinsic error rate which limits the visibility. (As we discuss later, a third source of errors also contributes to the QBER.) These are, firstly, the dark count noise of the detector, and, secondly, stray light from the intense clock laser which is not fully filtered by the wavelength-division multiplexing (WDM) filter. The probability of an error count per clock cycle ($P_e$) was measured to be $8.5 \times 10^{-7}$. Of this, the probability of a detector dark count in the 3.5 ns gate was measured to be $3.2 \times 10^{-7}$. This shows that the contribution from stray light dominates over that due to the detector dark count, suggesting that the visibility (and QBER) could be improved by stronger filtering of the clock laser.

The dependence of the visibility on fiber length ($L$) can be modelled using the expression:

$$V = \frac{I_{max} - I_{min}}{I_{max} + I_{min}} = \frac{\mu 10^{-\alpha L/10} \eta_{Bob}}{\mu 10^{-\alpha L/10} \eta_{Bob} + 2P_e} \qquad (1)$$

in which $\mu$ is the average photon flux leaving Alice's apparatus, $10^{-\alpha L/10}$ represents the fiber attenuation across a fiber with length $L$ and $\eta_{Bob}$ the detection efficiency of Bob's setup. Using the measured values of $\mu$=0.1, $\eta_{Bob}$=0.045 (which includes both the finite detector efficiency and Bob's apparatus transmission loss of 5 dB) and the specified value of fiber attenuation $\alpha$=0.2 dB/km, the calculated visibility, shown as the solid line in Fig. 2, fits the experimental data reasonably well.

The QBER can be predicted from the quantum interference visibility. The QBER is defined as the ratio of the number of erroneous bits in the sifted raw key to the total number of the sifted bits. For BB84 protocol, the QBER $e$ can be written as

$$e = \frac{0.5 \cdot P_e}{0.5 \cdot \mu 10^{-\alpha L/10} \eta_{Bob} + P_e} \approx \frac{1-V}{2} \qquad (2)$$

If the QBER is less than 11%,[2] Alice and Bob can form a shared key with minimal information known to an eavesdropper, by classical error correction[3,4] and privacy amplification.[5] The QBER calculated with Eq.2 and $P_e=8.5\times10^{-7}$ is shown as the dashed line in Fig.3, suggesting secure key distribution over 120 km is possible.

To achieve this range, the phase of the interferometer and the photon polarisation must be stable during the key distribution. A drift in the phase of the interferometer, due to variations in the relative lengths of the two arms, could contribute directly to the QBER. By casing both Alice's and Bob's setups in enclosures to prevent air convection, we found the phase drift rate to be less than 0.05° per second, allowing key distribution to be performed over several minutes. Polarisation drift reduces the bit rate, but does not degrade the QBER provided that the signal rate is significantly higher than the intrinsic error rate. We found that the polarisation is stable for >30mins for a 122 km fiber link without any noticeable drop in the bit formation rate.

QKD was performed using the well known BB84 protocol.[6] Two separate computers control Alice and Bob's electronics, and exchange the classical information (about encoding bases and photon detection times) required for key formation over the Internet using the TCP/IP protocol. The measured QBER, shown in Fig. 3, remains virtually constant at around 3.3% for fiber lengths up to 65 km. For these fiber lengths, the contributions due to detector dark counts and stray light are less than 0.4%, as shown by the calculated result (dashed line in Fig.3). Imperfections of the interferometer also play a minor role, as the classical interference visibility is better than 99.9%. The dominant contribution to the QBER for short fibers derives from errors in the phase modulation, resulting from slight inaccuracies of the phase modulator biases, as well as phase drift during the experiment. Beyond 65 km, the QBER increases with fiber length, due to the erroneous counts caused by the detector dark counts and stray light. The simulation, including both modulation errors and erroneous counts, shown as the solid line in Fig. 3, fit the experimental data well. At 122 km, the QBER averaged over a 2-minute key transfer is typically 8.9%. This error rate is below the 11% limit,[2] thus allowing us to perform error correction and privacy amplification to form a shared key as described below.

The sifted bit rate, as shown in Fig. 4, decreases with increasing fiber length at a rate of ~0.21 dB/km, close to the specified value of standard single-mode telecom fiber. The average sifted raw bit rate is 3.4 kbits/s for 4.4km of fiber, falling to 23.4 and 9.2 bits/s for 101 and 122km, respectively. To form a secure key we first applied a Cascade error correction routine,[4] followed by Privacy Amplification, which compressed the reconciled key to a much shorter one[13] and thereby reduced the information known to Eve. The open symbols in Fig. 4 show the key formation rate after error correction and privacy amplification. At QBER $e$=3.3%, the key formation rate is around half of the sifted raw bit rate, while at $e$=8.9% the key formation rate falls to 4.6% of the sifted raw bit rate.

The net bit rate falls to zero when the QBER approaches the security limit of 11%.[2] Interpolation of the collected data suggests the current system will approach the security limit (ie. QBER<11%) with a fiber length of 130 km. By eliminating the modulation errors and with better rejection of the clock laser light, the range of the system could be extended to 165 km.

*Indeed, the visibility of the quantum interference (µ=0.1) is measured to be 86% for a 165.8km fibre link after removing stray light by electronically synchronising Bob's detector, suggesting a QBER of 7.0% and therefore successful QKD is achievable at this length.* Further increases would then rely upon improvements in single photon detection technology, the receiver's transmission and the key formation protocols.

We discuss now pulse splitting attacks, where Eve uses the multi-photon photon pulses inevitably generated by a laser diode to gain information about the key.[13,14] In an optimal attack, Eve replaces her link to Bob with a lossless channel and allows only the multi-photon pulses to pass to Bob after removing exactly one photon for measurement. Such an attack is not feasible using today's technology, but is important to consider for guaranteeing unconditional security. To be secure from this attack in the worst case, the bit rate measured by Bob must exceed the rate of multi-photon pulses generated by Alice.[14] This condition imposes a limit of ~50 km for the current system. Unconditional security may be achieved for longer fibers and with higher bit rates by replacing the attenuated pulsed laser diode in the current system with a true single photon source.[15] Free space QKD with a single photon source has recently been demonstrated.[16,17]

In summary, we have demonstrated quantum key distribution over 122 km of standard telecom fiber using the BB84 protocol. The QBER at 122 km was measured as 8.9%. The dominant contributions to the QBER were identified as arising from phase modulation errors, false counts due to stray clock laser photons and detector dark counts, indicating further improvements in the range of quantum cryptography may be possible.

**References:**

[1] See review, N. Gisin, G. Ribordy, W. Tittel, and H. Zbinden, Rev. Mod. Phys. **74**, 145(2002).

[2] N. Lütkenhaus, Phys. Rev. A **59**, 3301(1999).

[3] C. H. Bennett, F. Bessette, G. Brassard, L. Salvail, and J. Smolin, Journal of Cryptology **5**, 3(1992).

[4] G. Brassard and L Savail, Lect. Notes Comp. Sci. *765*, 410(1994).

[5] C. H. Bennett, G. Brassard, C. Crépeau, and U. M. Mauer, IEEE Trans. Inform. Theory **41**, 1915(1995).

[6] C. H. Bennett and G. Brassard, *Proceedings of the IEEE International Conference on Computers, Systems and Signal Processing 175-179*, Bangalore, India(1984).

[7] P. D. Townsend, J. G. Rarity, and P. R. Tapster, Electron. Lett. **29**, 634(1993).

[8] C. Marand and P. D. Townsend, Opt. Lett. **20**, 1695(1995).

[9] A. Muller, T. Herzog, B. Hutner, W. Tittel, H. Zbinden, and N. Gisin, Appl. Phys. Lett. **70**, 793 (1997).

[10] R. J. Hughes, G. L. Morgan, and C. G. Peterson, J. Mod. Opt. **47**, 533 (2000).

[11] D. Stucki, N. Gisin, O. Guinnard, G. Ribordy, and H. Zbinen, New J. Phys. **4**, 41(2002).

[12] T. Hasegawa, J. Abe, H. Ishizuka, M. Matsui, T. Nishioka, and S. Takeuchi, CLEO/QELS 2003, QTuB1, Baltimore, USA, 2003.

[13] N. Lütkenhaus, Phys. Rev. A **61**, 052304(2000).

[14] G. Brassard, N. Lütkenhaus, T. Mor, and B. C. Sanders, Phys. Rev. Lett. **85**, 1330(2000).

[15] Z. Yuan, B. E. Kardynal, R. M. Stevenson, A. J. Shields, C. J. Lobo, K. Cooper, N. S. Beattie, D. A. Ritchie, and M. Pepper, Science **295**, 102(2002).

[16] E. Waks, K. Inoue, C. Santori, D. Fattal, J. Vuckovic, G. S. Solomon, and Y. Yamamoto, Nature **420**, 762(2002).

[17] A. Beveratos, R. Brouri, T. Gacoin, A. Villing, J-P. Poizat, and P. Grangier, Phys. Rev. Lett. **89**, 187901(2002).

**Figure captions**

**Figure 1** A schematic diagram showing the fiber optic quantum key distribution (QKD) system using BB84 phase encoding. WDM: wavelength division multiplexer; FS: fiber stretcher; PBS: polarization beam combiner/splitter; PC: polarization controller. The delay loop causes an optical delay of 5.8 ns between the short and long arms in Alice or Bob's interferemeter. Standard telecom fibers (Corning SMF-28) are used in the link between Alice and Bob.

**Figure 2** Quantum interference visibility as a function of fiber length for a laser strength of 0.1 photons per clock cycle. The solid line plots the value calculated with Eq.1 and measured parameters. The inset shows the classical interference fringes recorded for a 122 km link.

**Figure 3** The measured QBER as a function of fiber length (solid symbols). The dashed line shows the QBER calculated using Eq.2 and the measured detector erroneous counts, while the solid line is contribution from both the detector erroneous counts and the modulation error.

**Figure 4** The sifted raw bit formation rate (solid symbols) and the key formation rate after error correction and privacy amplification (open symbols) for different fiber lengths.

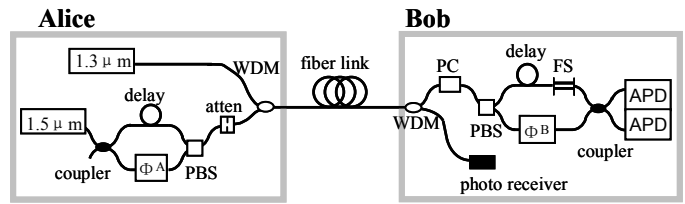

Fig. 1 Gobby et al

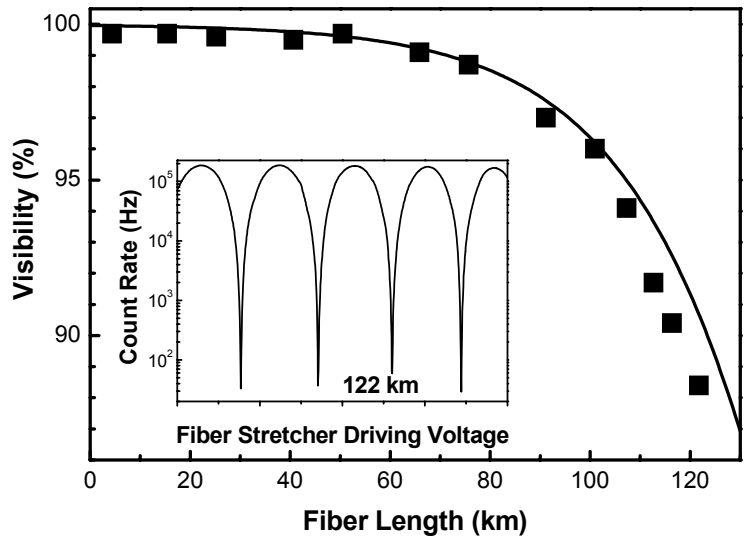

Fig. 2 Gobby et al.

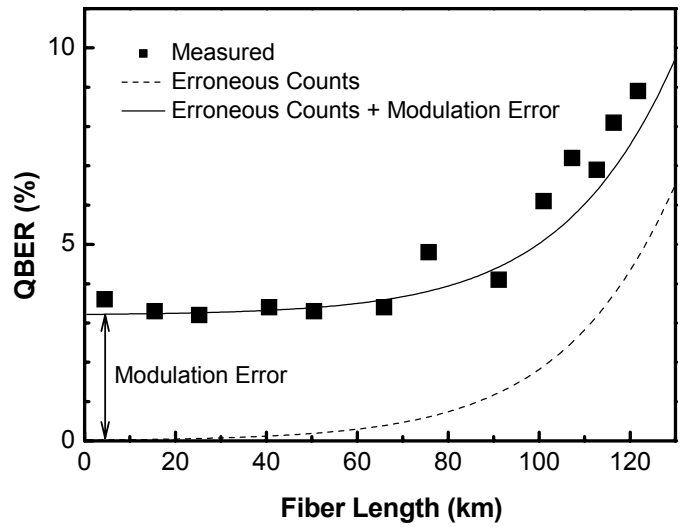

Fig. 3  Gobby et al.

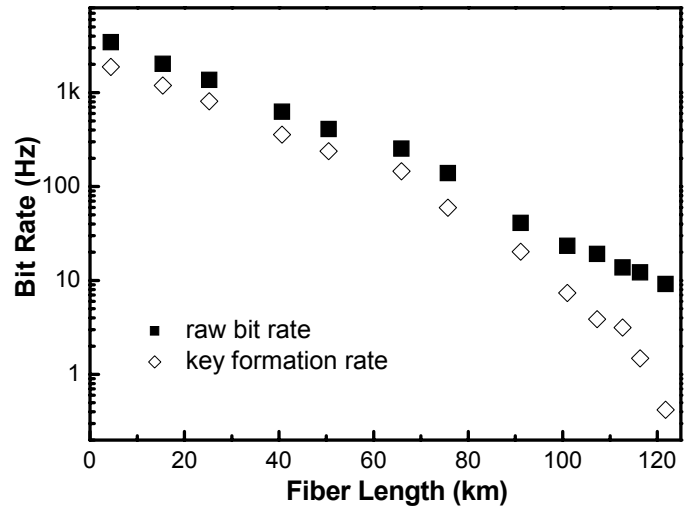

Fig. 4 Gobby et al.